\documentclass[11pt,english]{article}
\usepackage[T1]{fontenc}
\usepackage[latin1]{inputenc}
\usepackage{amsmath}
\usepackage{setspace}
\usepackage{amssymb}

\makeatletter
\newcommand{\lyxaddress}[1]{
\par {\raggedright #1
\vspace{1.4em}
\noindent\par}
}

\usepackage{babel}
\makeatother
\begin{document}

\title{QUATERNIONIC FORMULATION OF SUPERSYMMETRIC QUANTUM MECHANICS}

\author{Seema Rawat$^{\text{(1)}}$ and O. P. S. Negi$^{\text{(2)}}$}

\maketitle

\lyxaddress{\begin{center}$^{\text{(1)}}$Department of Physics\\
Govt. P.G.College\\
Ramnagar (Nainital), U.A.\par\end{center}}

\lyxaddress{\begin{center}$^{\text{(2)}}$Department of Physics\\
Kumaun University\\
S. S. J. Campus\\
Almora- 263601, U.A.\par\end{center}}

\lyxaddress{\begin{center}E-mail:- $^{\text{(1)}}$rawatseema1@rediffmail.com\\
$^{\text{(2)}}$ops\_negi@yahoo.co.in. \par\end{center}}

\begin{abstract}
Quaternionic formulation of supersymmetric quantum mechanics has been
developed consistently in terms of Hamiltonians, superpartner Hamiltonians,
and supercharges for free particle and interacting field in one and
three dimensions. Supercharges, superpartner Hamiltonians and energy
eigenvalues are discussed and it has been shown that the results are
consistent with the results of quantum mechanics. 
\end{abstract}

\section{ Introduction}

~~~~~Quaternionic quantum mechanics has been extensively studied
by Adler \cite{key-1}, while other authors \cite{key-2,key-3} revealed
out the noble features of quaternionic quantum mechanics. But the
subject has not been considered widely since there are various problems
with non-commutative nature of quaternion multiplication besides the
advance algebraic structure. On the other hand, supersymmetric quantum
mechanics is an application of SUSY superalgebra to quantum mechanics
as approved by quantum field theory. So one-dimensional SUSY has been
studied by various authors \cite{key-4,key-5}, and the efforts have
been made by various authors \cite{key-5,key-6,key-7,key-8,key-9,key-10,key-11}
to generalize it to higher dimensions. One of the attempts was also
made by Das et al \cite{key-2} to consider the higher dimensional
SUSY quantum mechanics. While the Cooper et al \cite{key-5} reviewed
the theoretical formulation of quantum mechanics and discussed many
problems therein. Supersymmetric quantum mechanics involves pairs
of Hamiltonians, which share a particular mathematical relationship,
which are called partner Hamiltonians and the potential energy terms
occur in Hamiltonians are then described as partner potentials. Accordingly,
for every eigenstate, of one Hamiltonian in partner Hamiltonian, has
a corresponding eigenstate with the same energy (except possible for
zero energy eigenstates). Each boson would have a fermionic partner
of eigen energy but in relativistic world energy and mass are interchangeable.
So one can say that partner particles have equal masses. SUSY concepts
have provided useful extension to WKB approximation \cite{key-6}.
Supersymmetric methods in quaternionic quantum mechanics are discussed
by Adler \cite{key-1} and Davies \cite{key-12} to study supersymmetric
quaternionic quantum mechanics.

Keeping in view the application of SUSY and quaternion quantum mechanics,
we have made an attempt in this paper to develop quaternionic quantum
mechanics from the basics of free particle quaternion differential
operator. Free particle superpartner Hamiltonian, supercharges and
total Hamiltonian are accordingly calculated. Introducing the interaction
through quaternion super potential, interacting super-partner Hamiltonians,
supercharges and total Hamiltonian are discussed consistently and
satisfies the properties of supersymmetric algebra. Because of non-commutative
nature of quaternions, we have made an attempt to solve the problem
by restricting the propagation along X-axis only and interacting operators
in one dimension are derived. Correspondingly, the supercharges, superpartner
Hamiltonians and total Hamiltonian are again discussed to satisfy
the SUSY algebra. It has been shown that the condition for good supersymmetry
is that supercharges must annihilate the vacuum state . Using this
condition, we have obtained ground state quaternionic wave function.
It has been shown that the quaternionic superpotential obtained in
this manner resembles with the result obtained earlier by Davies \cite{key-12}.With
the help of these operators Schrodinger wave equation is obtained
for Hamiltonian . Superpartner Hamiltonians are factorized in terms
of ceation and annihilation operators and and in that case our results
resemble with Sukumar \cite{key-10}. It has been shown that energy
eigenvalues of superpartner Hamiltonians are positive definite. The
ground state wave function has also been obtained in terms of quaternion
potentialand superpartner Hamiltonians are derived consistently. It
has also been shown that the second order superpotential describes
anti-commutation relations while the first order superpotential gives
rise to commutation relations of ceation and annihilation operators.
As such the first and second order superpotential describes respectively
the system of bosons and fermions. it has been calculated that the
energy eigenvalue of superpartner Hamiltonian is no vanising but equals
to the energy of first excited state. It has also been shown that
the energy of in first excited equals to energy of in second excited
state. We have also shown that the energy spectrum is related as energy
eigenstates are equally spaced. Our results are same as those obtained
earlier by Sukumar \cite{key-10} and Rajput \cite{key-13} and we
may conclude that quaternionic supersymmetric quantum mechanics is
consistent with supersymmetric quantum mechanics.

\section{Definition}

A quaternion $\phi$ is expressed as

\begin{eqnarray}
\phi & = & e_{\text{0}}\phi_{0}+e_{1}\phi_{1}+e_{2}\phi_{2}+e_{3}\phi_{3}\label{eq:1}\end{eqnarray}
Where $\phi_{0},\phi_{1},\phi_{2},\phi_{3}$ are the real quartets
of a quaternion and $e_{0},e_{1},e_{2},e_{3}$ are called quaternion
units and satisfies the following relations,

\begin{eqnarray}
e_{0}^{2} & = & e_{0}=1,\nonumber \\
e_{0}e_{i} & = & e_{i}e_{0}=e_{i}(i=1,2,3),\nonumber \\
e_{i}e_{j} & = & -\delta_{ij}+\varepsilon_{ijk}e_{k}(i,j,k=1,2,3),\label{eq:2}\end{eqnarray}
where $\delta_{ij}$ is the Kronecker delta and $\varepsilon_{ijk}$
is the three index Levi- Civita symbols with their usual definitions.
The quaternion conjugate $\bar{\phi}$ is then defined as \begin{eqnarray}
\bar{\phi} & = & e_{\text{0}}\phi_{0}-e_{1}\phi_{1}-e_{2}\phi_{2}-e_{3}\phi_{3}\label{eq:3}\end{eqnarray}
Here $\phi_{0}$is real part of the quaternion defined as\begin{align}
\phi_{0} & =Re\,\,\phi=\frac{1}{2}(\bar{\phi}+\phi)\label{eq:4}\end{align}
If $Re\,\,\phi=\phi_{0}=0$ , then $\phi=-\bar{\phi}$ and imaginary
$\phi$ is called pure quaternion and is written as

\begin{eqnarray}
Im\,\,\phi & = & e_{1}\phi_{1}+e_{2}\phi_{2}+e_{3}\phi_{3}\label{eq:5}\end{eqnarray}
The norm of a quaternion is expressed as 

\begin{eqnarray}
N(\phi) & = & \bar{\phi}\phi=\phi\bar{\phi}=\phi_{0}^{2}+\phi_{1}^{2}+\phi_{2}^{2}+\phi_{3}^{2}\geq0\label{eq:6}\end{eqnarray}
and the inverse of a quaternion is described as

\begin{eqnarray}
\phi^{-1} & = & \frac{\bar{\phi}}{\left|\phi\right|}\label{eq:7}\end{eqnarray}
While the quaternion conjugation satisfies the following property

\begin{eqnarray}
\overline{(\phi_{1}\phi_{2})} & = & \bar{\phi_{2}}\bar{\phi}_{1}\label{eq:8}\end{eqnarray}
The norm of the quaternion (\ref{eq:6}) is positive definite and
enjoys the composition law

\begin{eqnarray}
N(\phi_{1}\phi_{2}) & = & N(\phi_{1})N(\phi_{2})\label{eq:9}\end{eqnarray}
Quaternion (\ref{eq:1}) is also written as $\phi=(\phi_{0},\overrightarrow{\phi})$
where $\overrightarrow{\phi}=$$e_{1}$$\phi_{1}+e_{2}\phi_{2}+e_{3}$$\phi_{3}$
is its vector part and $\phi_{0}$is its scalar part. The sum and
product of two quaternions are

\begin{eqnarray}
(\alpha_{0},\overrightarrow{\alpha})+(\beta_{0},\overrightarrow{\beta}) & = & (\alpha_{0}+\beta_{0},\overrightarrow{\alpha}+\overrightarrow{\beta})\nonumber \\
(\alpha_{0},\overrightarrow{\alpha})\,(\beta_{0},\overrightarrow{\beta}) & = & (\alpha_{0}\beta_{0}-\overrightarrow{\alpha}.\overrightarrow{\beta},\,\alpha_{0}\overrightarrow{\beta}+\beta_{0}\overrightarrow{\alpha}+\overrightarrow{\alpha}\times\overrightarrow{\beta})\label{eq:10}\end{eqnarray}
Quaternion elements are non-Abelian in nature and thus represent a
division ring.

\section{Quaternion SUSY for Free Particle }

Let us define four differential operator as quaternion in the following
manner (on using natural units $c=\hbar=1$ and $i=\sqrt{-1}$ through
out the text);\begin{eqnarray}
\boxdot & = & e_{1}\partial_{1}+e_{2}\partial_{2}+e_{3}\partial_{3}+\partial_{4}=-i\,\frac{\partial}{\partial t}+e_{1}\frac{\partial}{\partial x_{1}}+e_{2}\frac{\partial}{\partial x_{2}}+e_{3}\frac{\partial}{\partial x_{3}}.\label{eq:11}\end{eqnarray}
The quaternion conjugate of this equation is described as

\begin{eqnarray}
\overline{\boxdot} & = & -e_{1}\partial_{1}-e_{2}\partial_{2}-e_{3}\partial_{3}+\partial_{4}=-i\,\frac{\partial}{\partial t}-e_{1}\frac{\partial}{\partial x_{1}}-e_{2}\frac{\partial}{\partial x_{2}}-e_{3}\frac{\partial}{\partial x_{3}}.\label{eq:12}\end{eqnarray}
Using the quaternion multiplication rule (\ref{eq:2}) we may write
the norm of the quaternion differential operator given bt equations
(\ref{eq:11}-\ref{eq:12}) as 

\begin{eqnarray}
N(\boxdot)=\boxdot\overline{\boxdot}=\overline{\boxdot}\boxdot & = & \partial_{1}^{2}+\partial_{2}^{2}+\partial_{3}^{2}+\partial_{4}^{2}=\frac{\partial^{2}}{\partial x_{1}^{2}}+\frac{\partial^{2}}{\partial x_{2}^{2}}+\frac{\partial^{2}}{\partial x_{3}^{2}}-\frac{\partial^{2}}{\partial t^{2}}\label{eq:13}\end{eqnarray}
Equation ( \ref{eq:13}) can also be related with the D'Alembertian
operator in the fallowing manner i.e.

\begin{eqnarray}
\square & = & \boxdot\overline{\boxdot}=\overline{\boxdot}\boxdot=\frac{\partial^{2}}{\partial x_{1}^{2}}+\frac{\partial^{2}}{\partial x_{2}^{2}}+\frac{\partial^{2}}{\partial x_{3}^{2}}-\frac{\partial^{2}}{\partial t^{2}}=-\frac{\partial^{2}}{\partial t^{2}}+\triangledown^{2}.\label{eq:14}\end{eqnarray}
Let us consider the quaternion differential operator ( in three space
dimensions) and describe it as the free particle operator i.e.

\begin{eqnarray}
\widehat{A}_{free}=\bigtriangleup & = & e_{1}\partial_{1}+e_{2}\partial_{2}+e_{3}\partial_{3}=\sum_{j=1}^{3}e_{j}\bigtriangledown_{j}\,\,\,\,\,\,\,(\bigtriangledown_{j}\,=\frac{\partial}{\partial x_{j}})\label{eq:15}\end{eqnarray}
The conjugate of equation ( \ref{eq:15} ) is then be written as

\begin{eqnarray}
\widehat{A}_{free}^{\dagger}=\bigtriangleup^{\dagger} & = & -e_{1}\partial_{1}-e_{2}\partial_{2}-+e_{3}\partial_{3}=\sum_{j=1}^{3}e_{j}^{\dagger}\bigtriangledown_{j}^{\dagger}\label{eq:16}\end{eqnarray}

where $^{\dagger}$ corresponds to quaternionic conjugation. So that
superpartner of a free particle Hamiltonian can be formed as 

\begin{eqnarray}
\hat{H_{1}}=\hat{H_{-}}=\widehat{A}_{free}^{\dagger}\widehat{A}_{free} & = & \sum_{j=1}^{3}e_{j}^{\dagger}\bigtriangledown_{j}^{\dagger}\cdot\sum_{j=1}^{3}e_{j}\bigtriangledown_{j}\nonumber \\
\hat{H_{2}}=\hat{H_{+}}=\widehat{A}_{free}\widehat{A}_{free}^{\dagger} & = & \sum_{j=1}^{3}e_{j}\bigtriangledown_{j}\cdot\sum_{j=1}^{3}e_{j}^{\dagger}\bigtriangledown_{j}^{\dagger}\label{eq:17}\end{eqnarray}
Accordingly, the supercharges are described as

\begin{eqnarray}
\hat{Q} & = & \left[\begin{array}{cc}
0 & \widehat{A}_{free}\\
0 & 0\end{array}\right]=\left[\begin{array}{cc}
0 & \sum_{j=1}^{3}e_{j}\bigtriangledown_{j}\\
0 & 0\end{array}\right]\nonumber \\
\hat{Q}^{\dagger} & = & \left[\begin{array}{cc}
0 & 0\\
\widehat{A}_{free}^{\dagger} & 0\end{array}\right]=\left[\begin{array}{cc}
0 & 0\\
\sum_{j=1}^{3}e_{j}\bigtriangledown_{j} & 0\end{array}\right]\label{eq:18}\end{eqnarray}
So the free particle Hamiltonian in 3-dimensions is described as 

\begin{eqnarray}
\hat{H}_{free} & =\widehat{H} & =\left\{ \begin{array}{cc}
\sum_{j=1}^{3}e_{j}\bigtriangledown_{j}\cdot\sum_{j=1}^{3}e_{j}^{\dagger}\bigtriangledown_{j}^{\dagger} & 0\\
0 & \sum_{j=1}^{3}e_{j}^{\dagger}\bigtriangledown_{j}^{\dagger}\cdot\sum_{j=1}^{3}e_{j}\bigtriangledown_{j}\end{array}\right\} .\label{eq:19}\end{eqnarray}

Here the supercharges (\ref{eq:18}) and Hamiltonian (\ref{eq:19})
satisfy the super symmetric algebra given by

\begin{eqnarray}
\left[\widehat{H}\,,\,\hat{Q}\right] & = & \left[\widehat{H}\,,\,\hat{Q}^{\dagger}\right]=0\nonumber \\
\left\{ \hat{Q,\,}\hat{Q}\right\}  & = & \left\{ \hat{Q}^{\dagger}\hat{,Q}^{\dagger}\right\} =0.\label{eq:20}\end{eqnarray}
Here equation (\ref{eq:20}) results in degeneracy of energy and hence,
the supercharges $\hat{Q}$ and$\hat{Q}^{\dagger}$and generate supersymmetric
transformations and change accordingly a bosonic state to a fermionic
state or vice versa. Relation $\widehat{H}=\left\{ \hat{Q,\,}\hat{Q}^{\dagger}\right\} $shows
that Hamiltonian can have only positive or zero eigen values i.e.

\begin{eqnarray*}
\left\langle \psi\left|\widehat{H}\right|\psi\right\rangle  & = & \left\langle \psi\left|\hat{Q\,}\hat{Q}^{\dagger}\right|\psi\right\rangle +\left\langle \psi\left|\hat{Q\,}\hat{Q}^{\dagger}\right|\psi\right\rangle \end{eqnarray*}
\begin{eqnarray}
= & \mid\hat{Q\,}\left|\psi\right\rangle ^{2}+\mid\hat{Q}^{\dagger}\left|\psi\right\rangle ^{2} & \geq0\label{eq:21}\end{eqnarray}

\section{Quaternion SUSY for Interacting Field}

~~~~Let us describe the interaction through the introduction of
quaternion super potential defined as 

\begin{eqnarray}
U & = & e_{1}U_{1}+e_{2}U_{2}+e_{3}U_{3}\label{eq:22}\end{eqnarray}
where $\overrightarrow{U}=(U_{1},U_{2},U_{3})$is the three dimensional
super potential. Then, the operators for this case of interaction
become

\begin{eqnarray}
\widehat{A} & = & \boxdot+U=\sum_{j=1}^{3}e_{j}(\bigtriangledown_{j}+U_{j})\nonumber \\
\widehat{A}^{\dagger} & = & \boxdot^{\dagger}+U^{\dagger}=\sum_{j=1}^{3}e_{j}^{\dagger}(\bigtriangledown_{j}^{\dagger}+U_{j}^{\dagger}).\label{eq:23}\end{eqnarray}
So that we may define the supercharges for interacting field as follows,

\begin{eqnarray}
\hat{Q} & = & \left[\begin{array}{cc}
0 & \widehat{A}\\
0 & 0\end{array}\right]=\left[\begin{array}{cc}
0 & \sum_{j=1}^{3}e_{j}(\bigtriangledown_{j}+U_{j})\\
0 & 0\end{array}\right]\nonumber \\
\hat{Q}^{\dagger} & = & \left[\begin{array}{cc}
0 & 0\\
\widehat{A}^{\dagger} & 0\end{array}\right]=\left[\begin{array}{cc}
0 & 0\\
\sum_{j=1}^{3}e_{j}^{\dagger}(\bigtriangledown_{j}^{\dagger}+U_{j}^{\dagger}) & 0\end{array}\right]\label{eq:24}\end{eqnarray}
while the super partner Hamiltonians are defined in the following
manner

\begin{eqnarray}
\hat{H_{1}}=\hat{H_{-}}=\widehat{A}^{\dagger}\widehat{A} & = & \sum_{j=1}^{3}e_{j}^{\dagger}(\bigtriangledown_{j}^{\dagger}+U_{j}^{\dagger})\cdot\sum_{j=1}^{3}e_{j}(\bigtriangledown_{j}+U_{j})\nonumber \\
\hat{H_{2}}=\hat{H_{+}}=\widehat{A}\widehat{A}^{\dagger} & = & \sum_{j=1}^{3}e_{j}(\bigtriangledown_{j}+U_{j})\cdot\sum_{j=1}^{3}e_{j}^{\dagger}(\bigtriangledown_{j}^{\dagger}+U_{j}^{\dagger}).\label{eq:25}\end{eqnarray}
So that Hamiltonian from equation is described as

\begin{equation}
\widehat{H}=\left\{ \hat{Q,\,}\hat{Q}^{\dagger}\right\} =\left[\begin{array}{cc}
\widehat{A}\widehat{A}^{\dagger} & 0\\
0 & \widehat{A}^{\dagger}\widehat{A}\end{array}\right]=\left[\begin{array}{cc}
H_{+} & 0\\
0 & H_{-}\end{array}\right]\label{eq:26}\end{equation}

or

\begin{equation}
\widehat{H}=\left\{ \begin{array}{cc}
\sum_{j=1}^{3}e_{j}(\bigtriangledown_{j}+U_{j})\cdot\sum_{j=1}^{3}e_{j}^{\dagger}(\bigtriangledown_{j}^{\dagger}+U_{j}^{\dagger}) & 0\\
0 & \sum_{j=1}^{3}e_{j}^{\dagger}(\bigtriangledown_{j}^{\dagger}+U_{j}^{\dagger})\cdot\sum_{j=1}^{3}e_{j}(\bigtriangledown_{j}+U_{j})\end{array}\right\} \label{eq:27}\end{equation}
As such, we can verify the algebra of Supersymmetry (SUSY) i.e.

\begin{eqnarray}
\left[\widehat{Q\,},\widehat{H}\right] & = & \left[\widehat{Q\,},\widehat{H}^{\dagger}\right]=0\nonumber \\
\left\{ \widehat{Q\,},\widehat{Q\,}\right\}  & = & \left\{ \widehat{Q\,}^{\dagger},\widehat{Q\,}^{\dagger}\right\} =0\nonumber \\
\left\{ \widehat{Q\,},\widehat{Q\,}^{\dagger}\right\}  & = & \widehat{H}\label{eq:28}\end{eqnarray}

which is same as that of equation (\ref{eq:20}).As such, the SUSY
is satisfied for the case of interacting field for which the quaternionic
formulation of supercharges and Hamiltonian are described by equations
(\ref{eq:24}) and (\ref{eq:27}).

Let us restrict the propagation along one dimension (say X- axis only)
and letting Y = Z = 0, for simplification, and choosing quaternionic
unit $e_{2}$ along x-axis. Then the annihilation and creation operators
respectively given by $\widehat{A}^{\dagger}$and $\widehat{A}$ are
derived as

\begin{eqnarray}
\widehat{A} & = & e_{2}\frac{d}{dx}+\widehat{U}(x)\nonumber \\
\widehat{A}^{\dagger} & = & e_{2}\frac{d}{dx}-\widehat{U}(x)\label{eq:29}\end{eqnarray}
So that supercharges are obtained as

\begin{eqnarray}
\hat{Q} & = & \left[\begin{array}{cc}
0 & \widehat{A}\\
0 & 0\end{array}\right]=\left[\begin{array}{cc}
0 & e_{2}\frac{d}{dx}+\widehat{U}(x)\\
0 & 0\end{array}\right]\nonumber \\
\hat{Q}^{\dagger} & = & \left[\begin{array}{cc}
0 & 0\\
\widehat{A}^{\dagger} & 0\end{array}\right]=\left[\begin{array}{cc}
0 & 0\\
e_{2}\frac{d}{dx}-\widehat{U}(x) & 0\end{array}\right]\label{eq:30}\end{eqnarray}
and accordingly we may write the super partner Hamiltonians as

\begin{eqnarray}
\hat{H_{1}}=\hat{H_{-}}=\widehat{A}^{\dagger}\widehat{A} & = & -\frac{d^{2}}{dx^{2}}+e_{2}\widehat{U\,}'(x)-\widehat{U}^{2}(x)\nonumber \\
\hat{H_{2}}=\hat{H_{-}}=\widehat{A} & \widehat{A}^{\dagger}= & -\frac{d^{2}}{dx^{2}}-e_{2}\widehat{U\,}'(x)-\widehat{U}^{2}(x)\label{eq:31}\end{eqnarray}
Hence the total Hamiltonian in one dimension reduces to the following
expressions

\begin{eqnarray}
\widehat{H} & = & \left[\begin{array}{cc}
-\frac{d^{2}}{dx^{2}}-e_{2}\widehat{U\,}'(x)-\widehat{U}^{2}(x) & 0\\
0 & -\frac{d^{2}}{dx^{2}}+e_{2}\widehat{U\,}'(x)-\widehat{U}^{2}(x)\end{array}\right]\label{eq:32}\end{eqnarray}
This Hamiltonian Hermitian i.e. $\widehat{H}\,\,=\widehat{H}^{\dagger}$and
its eigen values are real contrary to the quaternion quantum mechanics
\cite{key-1}. We may now relate the real and quaternion Hamiltonian
\cite{key-1} in the following manner 

\begin{eqnarray}
\widehat{H} & = & -e_{2}\widetilde{H}=i\,\widetilde{H}.\label{eq:33}\end{eqnarray}
Since $e_{2}$ has eigenvalue $\pm i$. Here $\widetilde{H}$ is the
quaternionic Hamiltonian defined \cite{key-1,key-12} as 

\begin{eqnarray}
\widetilde{H} & = & \left[\begin{array}{cc}
-e_{2}\frac{d^{2}}{dx^{2}}+\widehat{U}\,'(x)-e_{2}\widehat{U}^{2}(x) & 0\\
0 & -e_{2}\frac{d^{2}}{dx^{2}}-\widehat{U\,}'(x)-e_{2}\widehat{U}^{2}(x)\end{array}\right].\label{eq:34}\end{eqnarray}
Equation (\ref{eq:31}) may then now be written as\begin{eqnarray}
\hat{H_{1}}=\hat{H_{-}}=\widehat{A}^{\dagger}\widehat{A} & = & -\frac{d^{2}}{dx^{2}}+\widehat{V}_{-}(x)=-\frac{d^{2}}{dx^{2}}+\widehat{V}_{1}(x)\nonumber \\
\hat{H_{2}}=\hat{H_{+}}=\widehat{A} & \widehat{A}^{\dagger}= & -\frac{d^{2}}{dx^{2}}+\widehat{V}_{+}(x)=-\frac{d^{2}}{dx^{2}}+\widehat{V}_{2}(x)\label{eq:35}\end{eqnarray}
where $\widehat{V}_{1}(x)$ or $\widehat{V}_{-}(x)$ and $\widehat{V}_{2}(x)$
or $\widehat{V}_{+}(x)$ are known as super partner potentials and
are thus related to quaternionic potential $U$ in the following manner,

\begin{eqnarray}
\widehat{V}_{1}(x) & = & -e_{2}\widehat{U\,}'(x)-\widehat{U}^{2}(x)\nonumber \\
\widehat{V}_{2}(x) & = & e_{2}\widehat{U\,}'(x)-\widehat{U}^{2}(x).\label{eq:36}\end{eqnarray}
Here also we may establish the condition for good supersymmetry which
is known as unbroken supersymmetry and where the the supercharges
annihilate the vacuum i.e.

\begin{eqnarray}
\widehat{Q}\left|\psi_{0}\right\rangle  & = & \widehat{Q}^{\dagger}\left|\psi_{0}\right\rangle =0\label{eq:37}\end{eqnarray}
where the ground state wave function is defined in terms of two component
wave function $\left|\psi_{0}\right\rangle =\left[\begin{array}{c}
\psi_{a}(x)\\
\psi_{b}(x)\end{array}\right]$ along with $\psi_{a}(x)$ and $\psi_{b}(x)$ are again described
in terms of two component wave function of a quaternion in sympletic
representation i.e. 

\begin{eqnarray}
\psi_{a}(x) & = & \psi_{0}+e_{1}\psi_{1}\nonumber \\
\psi_{b}(x) & = & \psi_{2}-e_{1}\psi_{3}.\label{eq:38}\end{eqnarray}
Using equations (\ref{eq:30}), (\ref{eq:37}) and (\ref{eq:38}),
we get

\begin{eqnarray}
\left[\begin{array}{cc}
0 & e_{2}\frac{d}{dx}+\widehat{U}(x)\\
0 & 0\end{array}\right]\left[\begin{array}{c}
\psi_{a}(x)\\
\psi_{b}(x)\end{array}\right] & = & e_{2}\psi_{b}'(x)+\widehat{U}(x)\psi_{b}(x)=0\nonumber \\
\left[\begin{array}{cc}
0 & 0\\
e_{2}\frac{d}{dx}-\widehat{U}(x) & 0\end{array}\right]\left[\begin{array}{c}
\psi_{a}(x)\\
\psi_{b}(x)\end{array}\right] & = & e_{2}\psi_{a}'(x)+\widehat{U}(x)\psi_{a}(x)=0\label{eq:39}\end{eqnarray}
which leads to the following sets of equations i.e.\begin{eqnarray}
\widehat{U}(x) & = & -\frac{\psi_{a,b}'(x)e_{2}\psi_{a,b}^{\star}(x)}{\left|\psi_{a,b}(x)\right|^{2}}=\pm\frac{\psi'(x)e_{2}\psi^{\star}(x)}{\left|\psi(x)\right|^{2}}.\label{eq:40}\end{eqnarray}
Since $e_{2}$ has eigenvalues $\pm i$ . Replacing $e_{2}$ by $\pm i$
our theory gives rise to the results obtained by Davies \cite{key-12}
and accordingly, we may obtain the hierarchy of Hamiltonians or a
series of Hamiltonians $\widehat{H}_{1},\widehat{H}_{2},\widehat{H}_{3}..........\widehat{H}_{n}$.

Now, we may write Schrödinger's equation for $\widehat{H}_{1}(\widehat{H}_{-})$
as 

\begin{eqnarray}
\widehat{H}_{1}\psi_{0}^{(1)}(x) & =- & \frac{d^{2}\psi_{0}^{(1)}(x)}{dx^{2}}+\widehat{V}_{-}(x)\psi_{0}^{(1)}(x)\,\,\,\,\,(\hbar=2m=1)\label{eq:41}\end{eqnarray}
where 

\begin{eqnarray}
\widehat{H}_{1} & =\widehat{H}_{-}= & \widehat{A}_{1}^{\dagger}\widehat{A}_{1}=-\frac{d^{2}}{dx^{2}}+\widehat{V}_{-}(x)=-\frac{d^{2}}{dx^{2}}+\widehat{V}_{1}(x)\nonumber \\
\widehat{A}_{1} & = & e_{2}\frac{d}{dx}+\widehat{U}_{1}(x)\nonumber \\
\widehat{A}_{1}^{\dagger} & = & e_{2}\frac{d}{dx}-\widehat{U}_{1}(x)\label{eq:42}\end{eqnarray}

Similarly, we get Schrödinger's equation for $\widehat{H}_{2}(\widehat{H}_{+})$
as 

\begin{eqnarray}
\widehat{H}_{2}\psi_{0}^{(1)}(x) & =- & \frac{d^{2}\psi_{0}^{(1)}(x)}{dx^{2}}+\widehat{V}_{+}(x)\psi_{0}^{(1)}(x)\,\,\,\,\,(\hbar=2m=1)\nonumber \\
\widehat{H}_{2} & =\widehat{H}_{+}=\widehat{A}_{1} & \widehat{A}_{1}^{\dagger}=-\frac{d^{2}}{dx^{2}}+\widehat{V}_{+}(x)=-\frac{d^{2}}{dx^{2}}+\widehat{V}_{2}(x).\label{eq:43}\end{eqnarray}

As such, we may obtain the positive energy eigenvalues of both superpartner
Hamiltonians $\widehat{H}_{1}(\widehat{H}_{-})$ and $\widehat{H}_{2}(\widehat{H}_{+})$
and it is to be proved that operator $\widehat{A}$ converts the eigenstate
of $\widehat{H}_{1}(\widehat{H}_{-})$into the eigenstate of $\widehat{H}_{2}(\widehat{H}_{+})$.
Similarly operator $\widehat{A}^{\dagger}$converts eigenstate of
$\widehat{H}_{2}(\widehat{H}_{+})$ into eigenstate of $\widehat{H}_{1}(\widehat{H}_{-})$.
Thus we conclude that $\widehat{A}^{\dagger}$works as raising operator
and $\widehat{A}$ as lowering operator.

\section{Superpartner Hamiltonians For Quaternion Harmonic Oscillator}

~~~~~~Let us define the ground state wave function as

\begin{eqnarray}
\psi_{0}^{(-)} & = & C\,\exp(\int\overrightarrow{U}(s).d\overrightarrow{s})\label{eq:44}\end{eqnarray}
where $C$ is the normalization constant, $\overrightarrow{U}(s)$
is the quaternion potential and $d\overrightarrow{s}$ is quaternion
difference operator so that 

\begin{eqnarray}
\overrightarrow{U}(s) & = & \omega(e_{1}x_{1}+e_{2}x_{2}+e_{3}x_{3})\label{eq:45}\end{eqnarray}
and $\omega$ is the frequency of the oscillator. Restricting the
propagation in one dimension only, we get

\begin{eqnarray}
\overrightarrow{U}(s) & = & \omega e_{2}x_{2}=\omega e_{2}x;\,\,\,\,\,\,\,\, d\overrightarrow{s}=e_{2}dx.\label{eq:46}\end{eqnarray}
Then equation (\ref{eq:45}) reduces to

\begin{eqnarray}
\psi_{0}^{(-)} & = & C\,\exp(-\int\omega x.dx)=C\,\exp(-\frac{\omega x^{2}}{2})\label{eq:47}\end{eqnarray}
Superpartner potentials are then be expressed as

\begin{eqnarray}
\widehat{V}_{-}(x) & =e_{2}\widehat{U\,}'(x)-\widehat{U}^{2}(x) & =-\omega+\omega^{2}x^{2}\nonumber \\
\widehat{V}_{-}(x) & =-e_{2}\widehat{U\,}'(x)-\widehat{U}^{2}(x) & =\omega+\omega^{2}x^{2}.\label{eq:48}\end{eqnarray}
and we get\begin{eqnarray}
\hat{H_{1}}=\hat{H_{-}} & = & -\frac{d^{2}}{dx^{2}}+\widehat{V}_{-}(x)=-\frac{d^{2}}{dx^{2}}-\omega+\omega^{2}x^{2}\nonumber \\
\hat{H_{2}}=\hat{H_{+}} & = & -\frac{d^{2}}{dx^{2}}+\widehat{V}_{+}(x)=-\frac{d^{2}}{dx^{2}}+\omega+\omega^{2}x^{2}.\label{eq:49}\end{eqnarray}
It may readily be proved that $\widehat{U}^{2}(x)$ is proportional
to anti commutation of annihilation operator $\widehat{A}$ and creation
operator $\widehat{A}^{\dagger}$while the first derivative $\widehat{U\,}'(x)$
is proportional to the commutation of annihilation operator $\widehat{A}$
and creation operator $\widehat{A}^{\dagger}$multiplied by quaternion
unit $e_{2}$. In the case of quaternionic quantum mechanics the Hamiltonians
$\widehat{H}_{+}$and $\hat{H_{-}}$ are superpartner Hamiltonians
i.e. for any eigen function $\psi_{0}^{(-)}$ of $\hat{H_{-}}$ with
the corresponding eigenvalues $E$, $\widehat{A}\psi_{0}^{(-)}$is
an eigen function of $\widehat{H}_{+}$ with the same eigenvalue.
Similarly, for any eigenfunction $\psi_{0}^{(+)}$ of $\widehat{H}_{+}$,
$\widehat{A}^{\dagger}\psi_{0}^{(+)}$is an eigenfunction of $\hat{H_{-}}$
with the same eigenvalue. We may now calculate the energy eigenvalue
spectrum of Quaternion Harmonic Oscillator from the basic definition
of supersymmetry. Using equations (\ref{eq:47}) and (\ref{eq:48})
we get

\begin{eqnarray}
\hat{H_{1}}\psi_{0}^{(-)} & =0 & =E_{0}^{(-)}\psi_{0}^{(-)}=E_{0}^{(1)}\psi_{0}^{(1)}\label{eq:50}\end{eqnarray}
which shows that energy eigenvalue of $\hat{H_{-}}$is zero. This
eigenvalue can be considered as ground state energy and is the same
as those obtained earlier\cite{key-14} for the case of quaternion
supersymmetric harmonic oscillator. Similarly, we may calculate the
energy of superpartner Hamiltonian $\widehat{H}_{+}$or $\hat{H_{2}}$
as

\begin{eqnarray}
\hat{H_{2}}\psi_{0}^{(-)} & =2\omega C\,\exp(-\frac{\omega x^{2}}{2}) & =2\omega\psi_{0}^{(-)}=E_{0}^{(2)}\psi_{0}^{(1)}\neq0\label{eq:51}\end{eqnarray}
which shows that

\begin{eqnarray}
E_{0}^{(2)} & = & E_{0}^{(+)}=2\omega.\label{eq:52}\end{eqnarray}
It shows that ground state energy of $\widehat{H}_{+}(\hat{H_{2})}$is
not zero. Accordingly we may calculate 

\begin{eqnarray}
E_{0}^{(+)} & = & E_{1}^{(-)}=2\omega\nonumber \\
E_{2}^{(-)} & = & E_{1}^{(+)}=4\omega\label{eq:53}\end{eqnarray}
and so on. In other words we may write the general relation between
$n^{th}$ and $(n+1)^{th}$energy levels in the following manner

\begin{eqnarray}
E_{n+1}^{(-)} & = & E_{n}^{(+)}.\label{eq:54}\end{eqnarray}
Hence, the energy spectrum is related as the relation between two
consecutive energy eigenstates which are equally spaced. Thus our
results are same as those obtained earlier by Sukumar {[}10].

\end{document}